\documentclass[showpacs,preprintnumbers,amsmath,amssymb]{revtex4}

\usepackage{dcolumn}
\usepackage{bm}

\usepackage{graphicx}
\newcommand{\MeV}{\textrm{MeV}}

\begin{document}


\title
{\bf Theoretical Evaluations of the Fission Cross Section of the 77 eV Isomer of $^{235}$U
 }
\author
{J. Eric Lynn and A.C. Hayes}
\affiliation
{Theoretical Division, Los Alamos National Laboratory
Los Alamos, NM 87545}

\date{\today}

\begin{abstract}

We have developed models of the fission barrier (barrier heights and
transition state spectra) that reproduce
reasonably well the measured fission cross section of $^{235}$U from
neutron energy  of 1 keV to 2 MeV. From these
models we have calculated the fission cross section of the 77 eV isomer of
$^{235}$U over the same energy range. We find
that the ratio of the isomer cross section to that of the ground state lies between
about 0.45 and 0.55 at low neutron
energies. The cross sections become approximately equal above 1 MeV.
The ratio of the neutron capture cross section
to the fission cross section for the isomer is 
predicted to be about a factor of 3 larger for the isomer than for the ground state of 
$^{235}$U at kev neutron energies. 
We have also calculated the cross section for the population of the
isomer by inelastic neutron scattering form the $^{235}$U ground state. 
We find that the isomer is
strongly populated, and for $E_n = 1 MeV$ the $(n,n'\gamma)$ 
cross section leading to the population of the isomer is 
of the order of 0.5 barn. 
Thus, neutron reaction network calculations
involving the uranium isotopes in a high neutron fluence
are likely to be affected by the 77 eV isomer of $^{235}$U. 
With these same models the fission 
cross sections of $^{233}$U and $^{237}$U can be reproduced
approximately using only minor adjustments to the barrier heights.
With
the significant lowering of the outer barrier that is expected for
the outer barrier the general behavior of the fission
cross section of $^{239}$Pu can also be reproduced.
\end{abstract}
\maketitle
\section{ INTRODUCTION}
In environments of very high neutron flux the ultimate yields from the chains of nuclear
reactions depend not only on the cross sections of nuclei in their ground states but also, to less or
greater degree, on the cross sections of excited states. In stars at high temperatures, the states
involved will depend on the Boltzmann distribution of excitation while in more transient situations,
such as nuclear explosions, the longer-lived isomeric states will play key roles. Long lived isomers
are difficult to populate electromagnetically in a hot dense plasma. However, they can be populated
strongly via neutron capture, and if the neutron energies are high enough, by inelastic neutron scattering.
An especially
interesting example is that of $^{235}$U, which has an isomer at only 77 eV excitation with a half-life of
26 m. The thermal neutron fission cross section of this isomer has been measured \cite{meas} and found to
be about twice the value of the nucleus in its ground state. This has led to the speculation
 that the
cross section may also be higher for fast neutrons thus enhancing the fission yield in a transient,
extremely high neutron flux. Equally, a lower cross section would diminish the yield. Since there is
no method available at present to measure the fast neutron cross section of the isomer, a theoretical
evaluation is required to answer this question.

The purpose of this paper is to perform detailed calculations of the fission cross sections for
the ground state and isomeric state of $^{235}$U. The ground state of
$^{235}$U has spin and parity 7/2$^-$,
while
the isomer is 1/2$^+$.
The latter is the same as the ground state of $^{239}$Pu, which has a gross fission
barrier height (relative to the neutron separation energy) very similar to that of $^{235}$U.
It is therefore
pertinent as part of this investigation to determine the main physics
distinguishing the magnitude
and energy dependence of the fission cross sections of $^{235}$U and $^{239}$Pu.

The task involves the assessment of the main features of the double-humped fission barrier 
(barrier heights and penetrability parameters) from available data relating to the fission of the 
compound nucleus $^{236}$U. These parameters are somewhat dependent on models of the transition 
states at the barrier peaks. Therefore, a range of models was considered, 
adjustments made to obtain 
reasonable agreement with the measured fission cross section of $^{235}$U,
 and calculations made of the 
corresponding cross section of the isomeric state. For the models investigated, 
the fission cross-
section of the isomer is calculated to be substantially lower (by about $50\%$) 
than that of the ground 
state over a significant part of the neutron energy range (0 to $\sim$ 0.5MeV).
This work is described in Section II.

Using analysis of the same kinds of data as for the uranium isotope, barrier parameters were 
established for plausible models of the barrier transition states of 
the compound nucleus $^{240}$Pu, and 
from these parameters the fission cross section of $^{239}$Pu was calculated. 
This was found to agree 
well with the measured cross section. 
Similar calculations were carried out for the cross sections of 
$^{233}$U and $^{237}$U. 
These odd mass neighbors of $^{235}$U are expected to have similar barrier properties, 
while $^{237}$U, like $^{239}$Pu, 
has the same spin and parity as the $^{235}$U isomer. These studies help give 
confidence that we have sound understanding of the barrier and transition 
state systematics of this 
whole group of nuclei. This work is described in Section III.

\section{MODELS AND CALCULATIONS OF THE FISSION CROSS-SECTION OF $^{235}$U}

\subsection{General remarks}

\subsubsection{ Cross-section theory}

At moderately high excitation energies (up to the order of 10 MeV) Hauser-Feshbach theory\cite{HF} 
 is used for calculating cross sections. 
In this, the cross section is separated into its components 
of total angular momentum and parity, and each component is proportional to a spin-weighting 
factor multiplied by the ratio of the product of transmission coefficients $T$ 
for the entrance and exit 
channels and the sum over transmission coefficients for all channels. 
In its more sophisticated form 
extra factors have to be included to account for the statistical fluctuations of the partial widths of 
the underlying compound nucleus levels:

\begin{equation}
\sigma_{CC'} = 2\pi^2 \frac{\lambda^2 \sum_{c(e)} g_e T_{c(e)} \sum_{c'(o)} T_{c'(o)} S_{c(e),c'(o)}}{\sum_{c''}T_{c''}}
\end {equation}

Labels $c(e)$ refer to the entrance channel $c$ in its different possible spin combinations, 
$c'(o)$ to the 
outgoing channel $c'$ in spin combinations o, $\lambda$ is the de Broglie wavelength divided by
$ 2\pi$ and 
$S_{c(e),c'(o)}$  is the fluctuation averaging factor. 
Dresner\cite{Dresner}  has derived a numerical integral that can be 
used for the evaluation of $S$ in the general case when one channel 
(or, in practice, a large group of 
channels such as those accounting for radiative capture) is constant for individual compound 
nucleus levels while for each of the remaining channels the statistical 
distribution function over 
levels is a member of the $\chi^2$ family.

For neutrons with energies up to about 1 MeV bombarding very long-lived and well-studied 
actinide target nuclei such as $^{235}$U, 
the transmission coefficients for the entrance neutron channel 
and individual inelastic channels can be calculated with considerable confidence. 
Alternatively, and this is the procedure adopted here, we can use
the experimental information on $s$-wave and $p$-wave neutron strength functions.
The radiation 
widths are known with reasonable accuracy in the slow neutron resonance region 
and can be extrapolated using simple 
statistical models of the radiation process. The chief difficulty in applying the Hauser-Feshbach 
theory to the fissionable nuclei lies in the nature of the fission process.

It is well-known that the fission barriers of the actinide group of nuclides are double-
humped in their functional dependence on deformation. This contrasts with the single hump or 
maximum in the potential energy given by the liquid drop model of fission. A double-humped 
barrier has many consequences on the fission cross section as a 
result of the subtle inter-play of the 
two maxima and makes analysis of the data a complicated and not always unambiguous one. It is 
not only in the one variable ("prolate" deformation towards elongation and division) that the 
potential energy departs from the classical liquid-drop form. 
It is well-established that at the outer 
barrier of the fission path the nucleus is unstable to 
octupole deformations and the saddle-point here 
is at a mass-asymmetric shape. 
There is also strong, though indirect, evidence that at the inner 
barrier the nucleus is unstable to axial deformations. 
These different shapes affect the energies of 
the transition states 
(also known as Bohr channels, after the introduction of the concept by A.Bohr \cite{Bohr}), 
 the states of collective and quasi-particle excitation in which the nucleus passes over the 
barrier saddle-point. Once the transition states are established, by either 
experimental evidence or 
hypothesis, the transmission coefficients for each barrier can be calculated, 
and from these we can 
calculate the overall fission transmission coefficient.

At excitation energies well above the barriers, the fission transmission coefficient, $T_F$, has 
the Strutinsky statistical form \cite{strutinsky} that simply relates the overall transmission to the separate 
transmission coefficients for crossing the inner and outer barriers, $T_A$, $T_B$, respectively.
\begin{equation}
T_F = \frac{T_A T_B}{T_A + T_B}
\end{equation}

The coefficients $T_A, T_B$ are given by the Bohr and Wheeler 
prescriptions of the sum of transition 
states \cite{wheeler} each multiplied by a barrier penetrability factor:
\begin{equation}
T_A = \sum_f \frac{1}{1 + exp[(V_A + E_{f,A} - E)/\hbar\omega_A]}
\end{equation}

(and similarly for $T_B$).
In Eq.4 the penetrability factor is the Hill-Wheeler formula \cite{hill} 
for a barrier with parabolic form 
equivalent to an inverted harmonic oscillator with circular frequency $\omega, E$ 
is the excitation energy, 
$V_A$ is the inner barrier height and the $E_{f,A}$ 
are the energies of the transition states $f$ above the inner 
barrier. A similar equation can be written for the outer barrier. 
It is possible to define transmission 
coefficients for individual transition states $f$:
\begin{equation}
T_f = \frac{T_A T_{B,f}}{(T_A + T_B)}
\end{equation}

These, with their appropriate quantum numbers, are important for calculating cross sections for 
more specific properties of the fission reaction, such as fission product angular distributions.

At lower energies the intermediate structure due to compound- nucleus-type levels (class-II 
states \cite{classii}) associated with the deformation of the secondary well between the inner and outer barrier 
peaks must be taken into account. By concentrating the fission 
strength into narrow energy regions 
these lower the average fission probability \cite{prob}. 
Also the effect of Porter-Thomas fluctuations both 
in the fine-structure compound-nucleus levels (class-I levels) 
and the class-II levels must be taken 
into account. These fluctuation effects are discussed in ref.\cite{lynn},
where analytic and numerical 
results have been established for a few limiting cases. 

In the present work  
fluctuation averaging has been studied
much more extensively. From the transmission coefficients $T_A, T_{B,f}$ the
mean coupling and fission widths
$<\Gamma_{II(c)}>, <\Gamma_{II(f)}>$, respectively, of the class-II levels are obtained:
\begin{eqnarray}
<\Gamma_{II(c)}> = D_{II}T_A/2\pi\\
<\Gamma_{II(f)}> = D_{II}T_{Bf}/2\pi
\end{eqnarray}
where $D_{II}$ is the mean class-II level spacing. Monte Carlo techniques have been used for selection
of the parameters (coupling widths, fission widths and individual spacings) of the
individual class-II levels from Porter-Thomas distributions and, using the select class-II
coupling width values, the coupling matrix elements with the class-I levels (for which 
spacings and reduced neutron widths were also selected using pseudo-random numbers) were
drawn from zero-mean Gaussian distributions. Solution of the eigenvalue problem then gave 
the parameters for the resonance fine structure, from which the detailed cross section
could be computed and averaged.

Comparison of these results with the formula \cite{prob} based on a uniform picket
fence model of the intermediate and fine structure gives the fluctuation averaging factor.
The uniform picket fence formula for the fission probability is
\begin{equation}
P_F = \frac{1}{[1 + R^2 + 2R coth(\pi(\Gamma_{II(c)} + \Gamma_{II(f)})/D_{II})]}^{1/2}
\end{equation}
where
\begin{equation}
R= \frac{\Gamma_I(\Gamma_{II(c)} + \Gamma_{II(f)})D_{II}}{\Gamma_{II(c)}\Gamma_{II(f)}D_I}
\end{equation}
and $\Gamma_I$ is the mean total width of the class-I levels. When the mean class-II width is much less
than the class-II level spacing, Eq. 8 gives a fission probability up to an order of magnitude lower
than the value deduced from the Hauser-Feshbach formula with the statistical expression, Eq. 5,
for the fission transmission coefficient.
Inclusion of the fluctuation averaging factor can reduce the fission probability by up to another
factor of three of more.

In the case of overlapping intermediate resonances ($T_A + T_B >> 1)$ Eq. 8 gives the statistical
result. Even in this case, however, when the intermediate structure is washed out the class-II
level fluctuations should be taken into account in the evaluation of the fission transmission
coefficients of Eq. 3 and 5. This is done here using the Dresner numerical integral technique
applied to Eq.5.
The individual transition state components of $T_B$ are governed by independent Porter-Thomas
distributions. Although the magnitude of $T_A$ is governed by the transition states across the inner
barrier, its fluctuation properties are governed by the degree of overlap of the class-II resonances.
The frequency distribution is assumed to be a member of the chi-squared family with $\nu$  degrees of
freedom. The value of $\nu$ is evaluated from a picket-fence model of the class-II states. For large $T_A$
the value of $\nu$  is $T_A/2$.

        The Monte Carlo method for calculating the overall fluctuation averaging factor in the
intermediate structure case is too time consuming to apply in a full calculation of the neutron cross-
sections. However, we find that if we apply the product of the separate fluctuation factors for the
fission transmission coefficient, Eq.5, and for the fine structure in the Hauser-Feshbach formula to
the intermediate structure fission probability, Eq.8, we obtain a quite good approximation to the
Monte Carlo result. It is this approximation that we use in general in calculating neutron cross-
sections, although we have frequently used the more exact Monte Carlo method for calculating the
fission probability below the neutron separation energy in the analysis of $(d,pf)$ and $(t,pf)$ data for
deduction of fission barrier heights.

\subsubsection{Fission barrier properties.}

The overall systematics of the fission barrier parameters of the actinides were established in 
a review by Bjornholm and Lynn \cite{lynn}. Since that work variations on the detailed parameters of 
specific nuclides have been published by other authors, 
but our understanding of the broad trends 
remains unchanged. Inner barrier heights (denoted by $V_A$) vary little over the range 
$Th$ to $Cf$ for a 
given parity class. 
For the double-even parity class the barrier height for the uranium isotopes and 
their neighbors is about 5.5 MeV. For even-odd (or odd-even) and double-odd nuclides the inner 
barrier is about 0.5 and 1 MeV higher, respectively. Outer barrier heights (denoted by $V_B$) 
vary 
strongly with proton number. 
For the uranium nuclides they are about the same as the inner barriers, 
whereas for plutonium they are about 0.5 MeV lower. 
From a gross point of view the overall barrier 
heights of the compound nuclei $^{236}$U and $^{240}$Pu are very similar, 
as are their neutron separation 
energies, 
leading to the simple expectation that the neutron-induced fission cross sections of $^{235}$U 
and $^{239}$Pu should be similar. 
In fact the different relative heights of the inner and outer barriers lead 
to considerable differences in the cross sections of the two nuclides.

The energies, total angular momenta and parities $I^\pi$ of 
the transition states are as important 
as the barrier heights in the fission process. 
These are largely extrapolated from the nuclear 
spectroscopy known for the ground state deformation. For double-even fissioning nuclides the 
lowest transition state, at both barriers, is, of course, the `ground state' at 
the barrier deformation, 
with excitation energy zero and $I^\pi = 0^+$. 
Built on this is a rotational band with $I^\pi = 2^+, 4^+, 6^+$ etc., 
with rotational moment of inertia $\Im$ 
inferred to be about twice, for the inner, and thrice, for the outer 
barrier, of that of the ground state. Thus, the transition state energies are

\begin{equation}
E _f =  E_I = I(I + 1)\hbar^2/2\Im                                                                   
\end{equation}

with respect to the barrier height. Above the `ground' transition state there is, 
in even nuclides, an 
energy gap, which could be significantly larger than 1 MeV, 
that is devoid of quasi-particle excitations. 
In this energy gap, however, it is expected that there will be 
collective vibrations each with its own 
rotational band. 
The beta-vibrations are the best known of these from nuclear spectroscopy, but, 
being vibrations in the prolate deformation variable, 
which becomes largely the fission degree of 
freedom, they do not enter into consideration of the transition states. 

Apart from the beta-vibrations, there is strong spectroscopic evidence for the gamma-
vibrations, which are vibrations about axial symmetry, with spin-projection along the prolate 
deformation symmetry axis and parity $K^\pi = 2^+$. 
The gamma-vibration energy is about 0.8 MeV in 
the actinides; its energy at the barrier deformations is unknown, 
and this is one of the quantities that 
is varied in the modeling process discussed below. 
In particular, if the nucleus is stable but soft to 
gamma deformation then the gamma vibration energy would be expected to be much lower than 0.8 
MeV. 
Another possibility is that the nucleus at the barrier 
is stable for a certain degree of non-axial 
symmetry, in which case extra bands for rotation about the major deformation axis occur.
These 
possibilities have to be considered especially for the inner barrier.

Odd-parity octupole vibrations have a special role among the transition states; they provide 
the principal means for odd-parity states of the compound nucleus to decay through fission.
 Nuclear 
spectroscopy provides evidence for two of these. At the lower energy, generally about 0.5 to 0.8 
MeV in the actinides, is the $K^\pi = 0^-$ vibration, 
the "mass asymmetry" vibration with rotational band 
members $I^\pi = 1^-,3^-,5^-$ etc. 
It is assumed to lie at about the same energy at the inner barrier, but 
probably much lower at the outer barrier where the saddle-point has a mass-asymmetric shape \cite{moller}, 
and the vibration is a low frequency reflection of the nuclear shape through the potential hill at zero 
octupole deformation. The higher energy vibration, often known as the "bending" vibration,
 has $K^\pi=1^-$ 
with rotational band members $I^\pi = 1^-,2^-,3^-$ etc.,
 and is usually found above 0.9 MeV at normal 
deformation. In most of our detailed modeling, 
the mass asymmetry vibration is assumed to be at 
0.7 MeV at the inner barrier, 0.1 MeV 
at the outer barrier, while the bending vibration is taken to be 
0.8 MeV and 0.6 MeV, respectively.

\subsubsection{Statistical representations of barrier transition states.}

The collective states described above are those expected in the energy gap before the 
appearance of the quasi-particle excitations that result from breaking the pairing energy. The energy 
gap is well-known in the spectra at normal deformation, and is a little greater than 1 MeV in the 
actinides. Above this energy the levels are normally described by a statistical level density function. 
The simplest form that is used is an exponential with constant temperature:
\begin{equation}
\rho_{A,B}(E,J) = C_{A,B} (2J+1)exp[-J(J+1)/2\sigma^2] exp[(E - V_{A,B})/\theta]                                   
\end{equation}
where $\rho(E,J)$ is the density of transition states with zero angular 
momentum and single parity at
excitation energy $E$ and total angular momentum $J$, $\sigma$ is a spin dispersion constant 
and $\theta$ is the 
temperature parameter. The subscripts $A,B$ label inner and outer barrier, respectively. 
At excitation 
energies of several MeV a Fermi-gas (independent particle) form is more appropriate. For such a 
composite model Cameron \cite{Cameron} has given tables of parameters that fit level density data. For the 
actinides these parameters have been readjusted in ref.\cite{lynn}. It is found that the temperature 
parameter is about 0.5 MeV. Spin dispersion constants are in the range of approximately 5 to 6.

For statistical representation of the transition states above the energy gap at the barrier 
deformations, theory suggests that the energy gaps are somewhat higher and the temperatures 
somewhat lower than those at stable deformation \cite{lynn}. The level density constants, CA,B, are also 
expected to be greater than at stable deformation by a factor dependent on the symmetry of the 
barrier shape\cite{shape} . The fission cross section at excitation energies considerably below the barrier 
energy gaps is affected by these ``continuum" transition states because of their high density and the 
Hill-Wheeler tunneling effect. The level density parameters can be adjusted in calculating the 
neutron-induced fission cross section up to about 2 MeV neutron energy.

\subsection{ Barrier parameters of $^{236}$U and calculated fission cross sections}

\subsubsection{ Axially symmetric inner barrier model}

	The $^{235}$U$(d,p)$ and $^{234}$U$(t,p)$ reactions can reach excitation energies 
in the compound nucleus 
well below the neutron separation energy ($S_n = 6.53 $ MeV for $^{236}$U), and thus can explore the fission 
probability well below the fission barrier. These reactions give the most direct information on 
barrier heights. Measurements of the $(d,pf)$ and $(t,pf)$ 
reactions have been made by Back et al. \cite{back1,back2}.
 Both these reactions excite compound nucleus states with a wide range of total 
angular momenta. 
The results of calculations of the relative cross sections for spin and parity are 
given in these references, and these have been used in our fits to the data. The error of measurement 
assessed in the above references includes a $20\%$ systematic error in magnitude. 
Therefore only the 
shape of the fission probability curve in the region of the barrier gives useful information. In this 
energy region only competition between fission and radiation has to be considered. 

The assumption of stiff axial symmetry at the inner barrier implies a high energy for the 
gamma phonon band transition states. We have assumed its value to be 0.8 MeV, similar to the 
observed value at normal deformation. With this transition state model, it is found that the fission 
probability data on the $(d,pf)$ reaction are quite well reproduced with inner and outer 
barrier parameters:\\
\begin{eqnarray}\nonumber
	V_A = 5.2 \MeV, \hbar\omega_A = 1.05 \MeV \\
\nonumber
	V_B = 5.7 \MeV, \hbar\omega_B = 0.6 \MeV 
\end{eqnarray}
The energy variation of the $(t,pf)$ data is also quite well-reproduced by these parameters
although the magnitude above the barrier is not in agreement. In this respect the
$(d,pf)$ and $(t,pf)$ data seem inconsistent.

Using these barrier parameters and the model of individual transition states described 
above, the statistical level density parameters can be adjusted to obtain reasonable agreement 
between calculation and the measured fission cross section of $^{235}$U up to about 1.2 MeV. This
is about the value of the energy gap in the target nucleus $^{235}$U, and the individual levels up
to this energy seem to be quite completely known, thus accounting almost fully for the expected
inelastic scattering.
Adjusted 
parameters are for the barrier ``continuum'' states are:
\[
	C_A = 0.20 \MeV^{-1}, C_B = 0.05 \MeV^{-1},  \theta_A =  \theta_B = 0.42 \MeV
\]
with an energy gap of 1.65 MeV above the inner barrier and 1.03 MeV above the outer barrier.

For neutron energies above 1.2 MeV we also need to describe the states of the residual
nucleus for inelastic scattering by means of a level density formula. If we retain the above
parameters for the barrier state density, we obtain by least squares fitting
\[
C_R = 0.194\pm .045 MeV^{-1}, \theta_R = 0.54\pm .05 MeV
\]
This is a considerably lower density than recommended in refs. \cite{Cameron,lynn}, namely $C_R = 0.9 MeV^{-1},
\theta_R = 0.5$ MeV. If we are to retain the parameters of ref.\cite{lynn} we must assume that the barrier densities
change at 2.5 MeV and 2.0 MeV above the inner and outer barrier respectively. Then the new
barrier density parameters for this higher energy region are
\[
C_A = 0.76 MeV^{-1}, C_B = 0.19 MeV^{-1},  \theta_A =  \theta_B = 0.40 MeV.
\]

The ground state spin and parity of $^{235}$U are $I^\pi = 7/2^-$. 
At low energies (up to a few tens of 
keV) s-wave neutron absorption is predominant. 
Compound nucleus states of spin and parity $J^\pi = 3^-
,4^-$ are formed. 
Transition states with these quantum numbers are fully open for both barriers at the 
neutron separation energy. At neutron energies of 50 keV p-wave neutron absorption has become 
comparable with s-wave absorption. The resulting compound nucleus states with $J^\pi$ 
ranging from $2^+$ 
to $5^+$ access transition states that are fully open over the inner barrier, 
the even spins are fully open 
over the outer barrier while the odd spins are about half-open there. The d-waves (exciting $J^\pi$ 
ranging from $1^-$ to $6^-$) become significant, 
but not dominant at about 0.5 MeV neutron energy. The 
corresponding transition states are essentially fully open. 
By contrast the 77eV isomer has spin $I^\pi = 
1/2^+$. The s-wave compound nucleus states have $J^\pi = 0^+,1^+$, 
the latter carrying three quarters of the 
compound nucleus formation cross section. 
The $J^\pi = 0^+$ transition state is open at the neutron 
separation energy for both barriers, but the more important $1^+$ state 
(a bending plus mass asymmetry 
combination) is about 0.3 MeV higher at the inner barrier and perhaps about equal to the neutron 
separation energy at the outer barrier. For this reason the low energy fission cross section of the 
isomer is calculated to be considerably lower than that of the ground state. The p-wave neutron 
absorption excites compound nucleus states with $J^\pi = 0^-,1^-$ and $2^-$. The lowest $0^-$ 
transition is not 
believed to exist within the energy gap (which is at approximately 7 MeV excitation for both 
barriers). Fission of the $0^-$ compound states 
(one twelfth of the compound nucleus formation cross-
section) is thus suppressed at low neutron energies.

The fission cross section calculated for the ground state from this barrier model, which we 
call Model 1, is shown in Figure 1, where the experimental data \cite{wasson, weston, poenitz} are also plotted. It is in fair 
agreement with the experimental data up to nearly 2MeV. The ratio of the fission cross section 
calculated for the isomer to the calculated cross section for the ground state is shown in Figure 2.

\subsubsection{ Axially asymmetric inner barrier: rigid rotator, $\gamma =11^0$}

	The degree of axial asymmetry of a rigid rotator is expressed by the conventional $\gamma$ 
parameter, in which $\gamma = 0$ describes a prolate spheroid, and $\gamma = 30^0$ describes maximum axial 
asymmetry. This model assumes $\gamma = 11^0$, a moderate degree of axial asymmetry. With the ground 
state rotational band inertial constant taken as $\hbar^2/2\Im = 3.33$ keV (giving the first $2^+$ 
rotational state 
at $\approx 20 $keV) the first $2^+, 3^+, 4^+$ etc. band (which can be thought of approximately as a gamma 
rotational band) occurs at about 250 keV, while a $4^+,5^+,6^+$ etc. band (`2 gamma' rotational) starts at 
about 1 MeV \cite{16}. Apart from higher bands involving combinations with the gamma bands, other 
transition states are similar to those in Model 1.

	The $^{235}$U$(d,pf)$ fission probability data are quite well reproduced with inner 
and outer barrier parameters:
\begin{eqnarray}\nonumber
	V_A = 5.53 \MeV, \hbar\omega_A = 1.05 \MeV \\
\nonumber
	V_B = 5.53 \MeV, \hbar\omega_B = 0.6 \MeV 
\end{eqnarray}
	Using these barrier parameters and the model of individual transition states described 
above to calculate the neutron fission cross section up to 1.2 MeV,
 the statistical level density parameters can be adjusted as in Model 1 to obtain:
\[
	C_A = 0.34\pm.07 \MeV^{-1},  C_B = 0.07\pm .02 \MeV^{-1},  \theta_A =  \theta_B = 0.475\pm .03 \MeV
\]
with an energy gap of 1.25 MeV above the inner barrier and 1.15 MeV above the outer barrier.
Above 1.2 MeV neutron energy, the residual nucleus level density is described with
parameters $C_R=0.212\pm .05 $ MeV, $\theta_R =0.566\pm .04$ MeV relative to the barrier state density.

The effect of Model 2 on the 77eV isomer is to lower the s-wave contribution to the cross-
section even more than in model 1.
The fission cross sections calculated with this barrier model are also given in Figures 1 and 
2.

\subsubsection{ Axially asymmetric inner barrier: rigid rotator, $\gamma =30^0$}

The ground state rotational band inertial constant is again taken as $\hbar^2/2\Im$ = 3.33 keV. The 
higher bands, based on $2^+$ and $4^+$ states deviate more from the rotational form, but the rotational 
relations can be used approximately with an inertial constant of 5.9 keV. The lower band starts at 
0.06 MeV and the higher band at 0.2 MeV. There is also a `3 gamma' band starting with $J^\pi =6^+$ at 
0.4 MeV.

The $^{235}$U$(d,pf)$ fission probability data are reproduced with the same inner and outer 
barrier parameters as in model 2.
Using these barrier parameters and the model of individual transition states described 
above, the statistical level density parameters can be adjusted to obtain agreement between the model 3 calculation
and the neutron fission cross section up to 1.2 MeV to  obtain: 
\[
	C_A = 0.34 \MeV^{-1}, C_B = 0.07 \MeV^{-1}, \theta_A = \theta_B = 0.46 \MeV
\]
with an energy gap of 1.32 MeV above the inner barrier and 1.22 MeV above the outer barrier.
Above 1.2 MeV the parameters of the residual nucleus level density are
$C_R =0.22\pm 0.05 MeV^{-1}, \theta_R = 0.504\pm 0.05 $MeV relative to the barrier state density.

The fission cross sections calculated with this barrier model are shown in Figures 1 and 2. 
The low energy fission cross section for the ground state is higher than in Model 2, but the ratio of 
the isomer and ground state  
cross sections is about the same.

\subsubsection{ Axially soft inner barrier}

The assumption that the nucleus at its inner barrier prolate deformation is soft to axially 
asymmetric distortions will lower the estimates of the gamma vibrational energy. We assume for 
our calculations with this model that the gamma phonon energy is 0.25 MeV and that these 
vibrations are harmonic. The rotational inertial constant is assumed to be 3.33 keV for all bands.

Again, the $^{235}$U$(d,pf)$ fission probability data can be reproduced with inner 
and outer barrier parameters as in model 2.

	Using these barrier parameters and this model of individual transition states, the statistical 
level density parameters can be adjusted as in previous models to obtain:
\[
	C_A = 0.34 \MeV^{-1}, C_B = 0.07 \MeV^{-1}, \theta_A = \theta_B = 0.47 \MeV
\]
with an energy gap of 1.1 MeV above the inner barrier and 1.15 MeV above the outer barrier.
The residual nucleus level density parameters $C_R = 0.19\pm0.05 MeV^{-1}, \theta_R =0.546\pm0.05$ MeV relative
to the barrier state density.

	The fission cross section calculated with this barrier model is shown in Figure 1, while the 
ratio of the isomer cross section to the ground state cross section is in Figure 2.  
 
Graphs of the cross sections of the
competing reactions are given in Figures 3 and 4 for Models 3 and 4, respectively.
All cross sections have been corrected for the  $(n,\gamma n')$ and $(n,\gamma f)$ reactions.
Some calculations have also been made of the cross section for populating the isomer
from the ground state by the 
$(n,n'\gamma)$ reaction. Different results are obtained depending on the assumption that K-quantum 
number selection rules apply to the cascading gamma-transitions or not. 
However, in all cases the isomer is predicted to be strongly populated via 
the $(n,n'\gamma)$ reaction on the $^{235}$U, and the cross section leading to the isomer at
neutron energies $\sim 1 MeV$ is of the order of 0.5 barn. 
These results are shown in 
Tables I and II.

\section{ CALCULATIONS OF THE CROSS-SECTIONS OF RELATED ACTINIDES}

\subsection{ The neutron-induced fission cross section of $^{233}$U}

	In this calculation we consider only Model 4 of Section II. The spin and parity of the target 
nucleus is $5/2^+$ and the neutron separation energy of the compound nucleus is 6.84 MeV. Fitting to 
the $^{233}$U$(d,pf)$ fission probability data suggests 
\begin{eqnarray}\nonumber
V_A &=& 5.83 MeV, \hbar\omega_A = 1.05 \MeV\\ 
\nonumber
V_B &=& 5.83 MeV, \hbar\omega_B = 0.7 \MeV 
\end{eqnarray}
Using these barrier parameters and transition state Model 4, and barrier state density
parameters that are close to those deduced for the $^{235}$U$(n,f)$ models 
($C_A =0.34 MeV^{-1}, C_B=0.157 MeV^{-1}, \theta_a=\theta_B= 0.45 MeV, C_R=0.23\pm.05 MeV^{-1}, 
\theta_R=0.465\pm.05 MeV$ relative to the barrier state density)
 we calculate neutron fission cross 
sections in good agreement with the data \cite{Meadows, Gwin}. We show the comparison in Figure 5.

\subsection{ The neutron-induced fission cross section of $^{237}$U}

	Again we consider only Model 4 of Section II. Like the isomeric state of $^{235}$U the spin and 
parity of the target nucleus is $1/2^+$ and the barrier transition states governing the fission cross-
section should therefore be very similar with the two nuclides differing by only two neutrons. 
However, the neutron separation energy of the compound nucleus is lower: 6.15 MeV. Fitting to the 
$^{236}$U$(t,pf)$ fission probability data suggests 
\begin{eqnarray}\nonumber
	V_A &=& 5.73 MeV, \hbar\omega_A = 1.05 MeV\\ 
\nonumber
	V_B &=& 5.83 MeV, \hbar\omega_B =  0.7 MeV 
\end{eqnarray}
Using these barrier parameters and transition state Model 4, we calculate neutron fission cross-
sections that are in fair agreement with the rather sparse data; these are limited to a single one-pulse time-of-
flight measurement on the Pommard shot \cite{Pommard} 
and a ratio measurement relative to the $^{235}$U fission 
cross section in a critical assembly \cite{crit}. We use this ratio as a normalization factor on the 
differential data. The comparison between calculation and adjusted data is shown in Figure 6.
The single point at 1.5 keV neutron energy is extrapolated from the resonance information measured in ref. \cite{Pommard}
 The 
generally lower trend of the calculation compared with the data suggests that  
the barrier heights are too high. However, to achieve agreement with the data in the 100 keV to 1 MeV
range calculations shows that the barriers would have to be reduced by about 200 keV, which is incompatible
with the analysis of the (t,pf) and resonance parameter data.

\subsection{ The neutron-induced fission cross section of $^{239}$Pu}

	Again we consider only Model 4 of Section II. Like the isomer of $^{235}$U, the spin and 
parity of $^{239}$Pu is $1/2^+$ and the relevant barrier transition states should be similar, 
as is the neutron 
separation energy of the compound nucleus: 6.53 MeV. However, the known general systematic 
trends of the double-humped fission barrier heights suggest that while the inner barrier height will 
be similar to that of the uranium nuclides we have studied above, the outer barrier may be about 0.5 
MeV lower. This is confirmed by fitting to the $^{239}$Pu$(d,pf)$ 
fission probability data, which agrees 
with 
\begin{eqnarray}\nonumber
	V_A &=& 5.63 MeV, \hbar\omega_A = 1.05 MeV\\ 
\nonumber
	V_B &=& 5.13 MeV, \hbar\omega_B = 0.7 MeV 
\end{eqnarray}
Using these barrier parameters and transition state Model 4, we calculate neutron fission cross-
sections that are in fairly good agreement with the data\cite{Gwin, Mclane}. The comparison is shown in Figure 7.
The dashed curve has been calculated with the same barrier state densities as in Model 4 of the $^{235}$U$(n,f)$ case.
For the bold curve the energy gaps at the barriers have been raised by 0.1 meV.
  The 
low energy fission cross section is considerably lower than that of the ground state of $^{235}$U 
(even though the fast neutron cross section is considerably higher) 
because of the high energy of the $1^+$ 
transition state at the inner 
barrier.  Above 0.1 MeV where the p-wave absorption predominates, the cross section is higher 
than that of $^{235}$U because of the considerably lower outer barrier.

\section{ CONCLUSIONS}

	We have derived double-humped fission barrier parameters and transition state spectra of 
the compound nucleus $^{236}$U that are consistent with known physics of the fission process and agree 
with data on fission probability extending into the sub-barrier energy region and with the neutron 
fission cross section of the ground state of $^{235}$U. With these fission barrier properties we have 
calculated the fission cross section of the 77eV isomer of $^{235}$U. The key transition state in this 
calculation is the $J^\pi = 1^+$ state. 
Both theoretical and experimental evidence suggest that this is at a 
high energy, especially at the inner barrier. This is the most important transition state for s-wave 
neutron induced fission, and therefore causes a considerable lowering of the low energy part of the 
cross section relative to the ground state cross section ($\sim 45-55\%$ in the models studied).  
The isomer cross section does not reach near-equality with the ground state cross section until the neutron 
energy is well above 0.5 MeV. The two cross sections then remain nearly equal until at least 2 MeV. 

The predicted ratio of the neutron capture cross section to the fission cross section
for the isomer is particularly striking. For model 3, for example, this ratio is predicted to
be a factor of about 3.4 (2.5) times larger than for the $^{235}$U ground state at 1 keV (10 keV). 
Figure 8 shows these predicted ratios, and the isomer capture to fission ratio remains larger up
to neutron energy of about 0.5 MeV
As discussed earlier and shown in Tables I and II,
the isomer is
strongly populated by inelastic neutron scattering on the $^{235}$U ground state.   
Thus, neutron reaction network calculations
involving the uranium isotopes in a high neutron fluence
are likely to be affected by the 77 eV isomer of $^{235}$U. 

	With one of these models (model 4), which is intermediate in properties amongst the set studied, 
the calculations of the fission cross sections are in good agreement with the 
measured cross sections for $^{233}$U, $^{237}$U and $^{239}$Pu. 
Some minor adjustments in barrier heights have 
to be made for the two uranium isotopes (to agree with the sub-barrier fission probability data). The 
agreement for $^{237}$U is most significant, because this has, like the $^{235}$U isomer, 
spin and parity $I^\pi = 
1/2+$. For the $^{239}$Pu case the outer barrier has to be reduced by about 
0.5 MeV to obtain agreement 
with the sub-barrier data, but with this change and the same transition state model all the major 
differences between the fission cross sections of $^{235}$U and $^{239}$Pu can be explained.

\newpage

\begin{table}
\caption{ Calculated cross sections for inelastic scattering from the ground state leading to 
population of the isomer and the ground state (the ground state cross section includes the 
compound elastic scattering). The assumption is made that all states below 2 MeV in $^{235}$U have 
good K-numbers.}

\begin{tabular}{ccc}

Neutron energy   &  $ \sigma(n,n'\gamma\rightarrow is)$ &   $\sigma(n,n'\rightarrow gd)$ \\
(MeV) &               (b)  &                  (b)\\  
\hline
	    0.1  &     0.21 &         0.86\\
            0.3 &      0.46 &         1.05\\   
	    0.5 &     0.51  &         0.97\\  
	    0.7 &     0.46  &        1.01 \\
	    0.9 &     0.42 &         1.01\\ 
	    1.1 &     0.40 &       0.96\\
	    1.3 &     0.54 &       0.87\\
	    1.5 &     0.77 &       0.75\\
	    1.7 &     0.96 &      0.66\\
\hline
\end{tabular}
\end{table}

\begin{table}
\caption{ Calculated cross sections for inelastic scattering from the ground state leading to 
population of the isomer and the ground state. The assumption is made that there is complete K  
mixing.}
\begin{tabular}{ccc}

Neutron energy &  $\sigma(n,n'\gamma\rightarrow is)$ & $\sigma(n,n'\gamma\rightarrow gs)$\\ 
 (MeV) &  (b) &   (b)\\  
\hline
	  0.1&                   0.21&                  0.86\\
	  0.3&                   0.63 &                 0.88\\
	  0.5&                   0.85&                  0.63\\
	  0.7&                   0.86&                  0.60 \\
	  0.9&                   0.87&                  0.56\\
	  1.1&                   0.83&                  0.53\\
	  1.3&                   0.83&                  0.58\\
	  1.5&                   0.87&                  0.65\\
	  1.7&                   0.91&                  0.72\\
\hline
\end{tabular}
\end{table}

\newpage

\begin{figure}
\vspace*{2cm}
\includegraphics{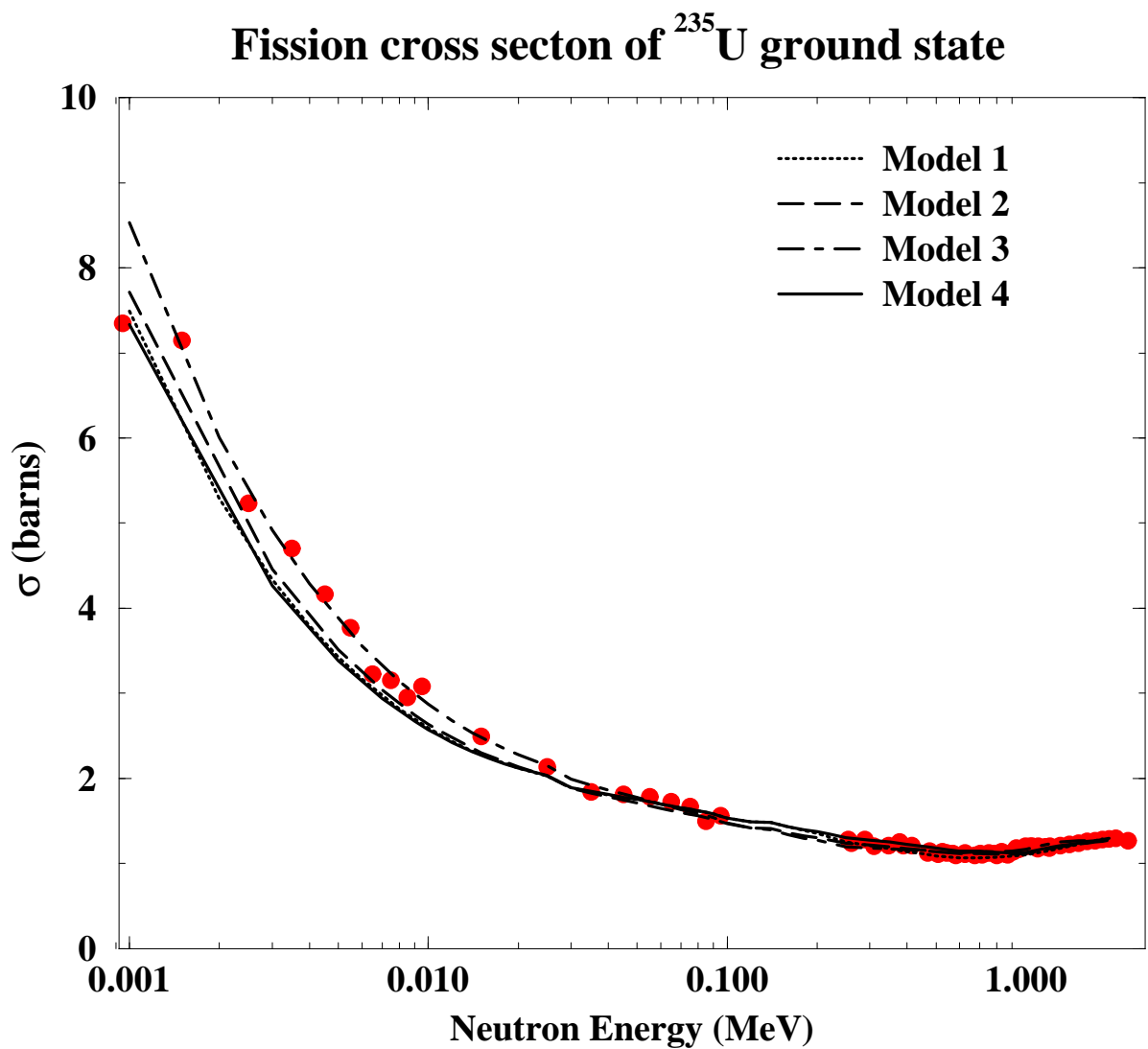}
\caption{The fission cross section of the $^{235}$U ground state for Models 1,2, and 3.
}
\end{figure}

\begin{figure}
\vspace*{2cm}
\includegraphics[width=5in]{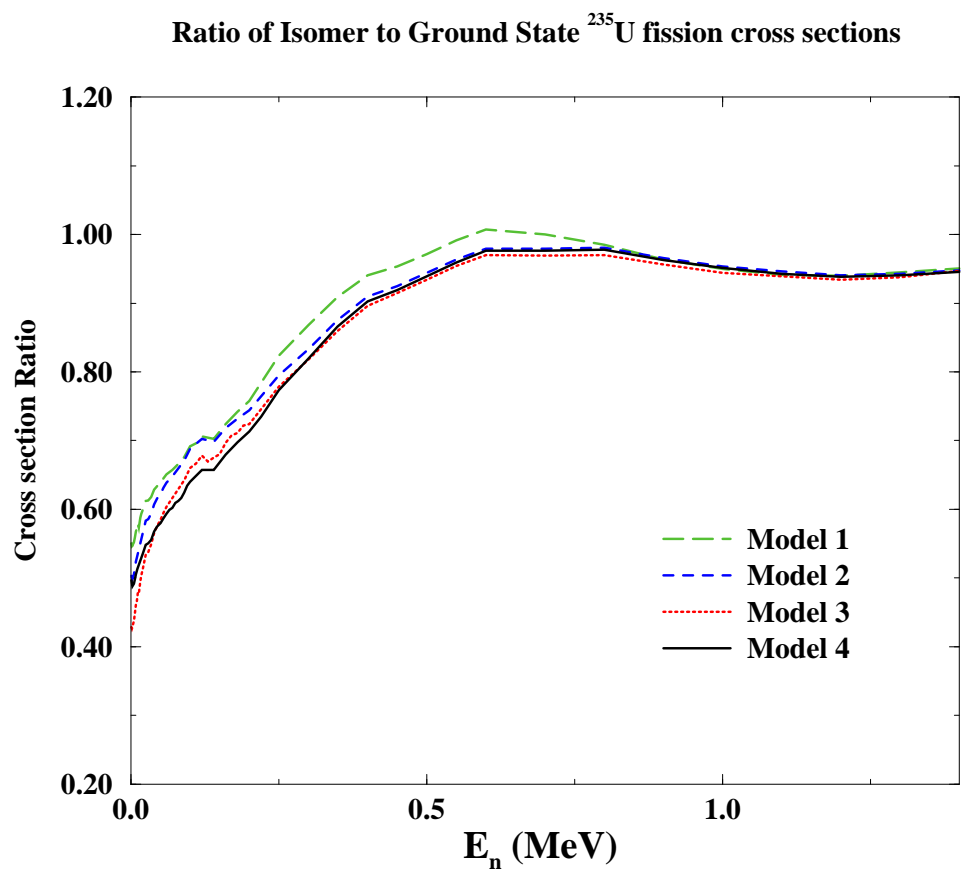}
\caption{ratio of calculated cross section for the 77 eV isomer of  $^{235}$U to
that of the ground state 
}
\end{figure}

\begin{figure}
\vspace*{2cm}
\includegraphics[width=5in]{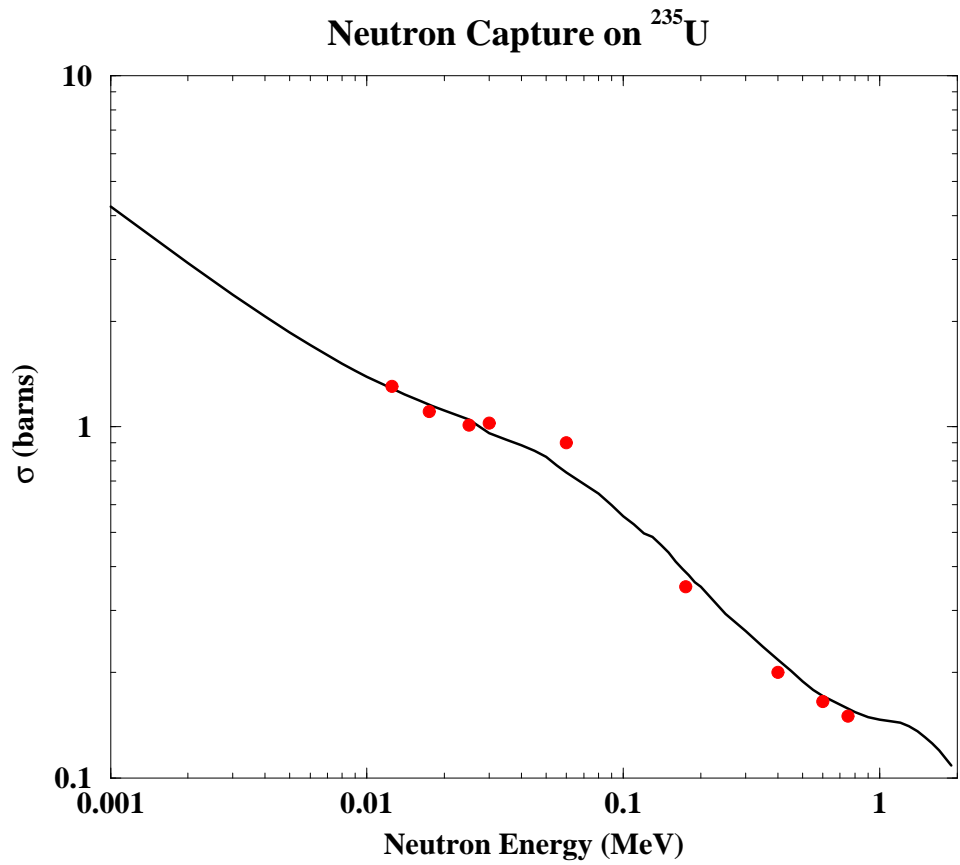}
\caption{The capture cross section of the $^{235}$U ground state for Models 4. 
As can be seen from Fig. 8, the capture cross section for the isomer is predicted to be
significantly larger than that for the ground state below 0.5 MeV.} 
\end{figure}

\begin{figure}
\vspace*{2cm}
\includegraphics[width=5in]{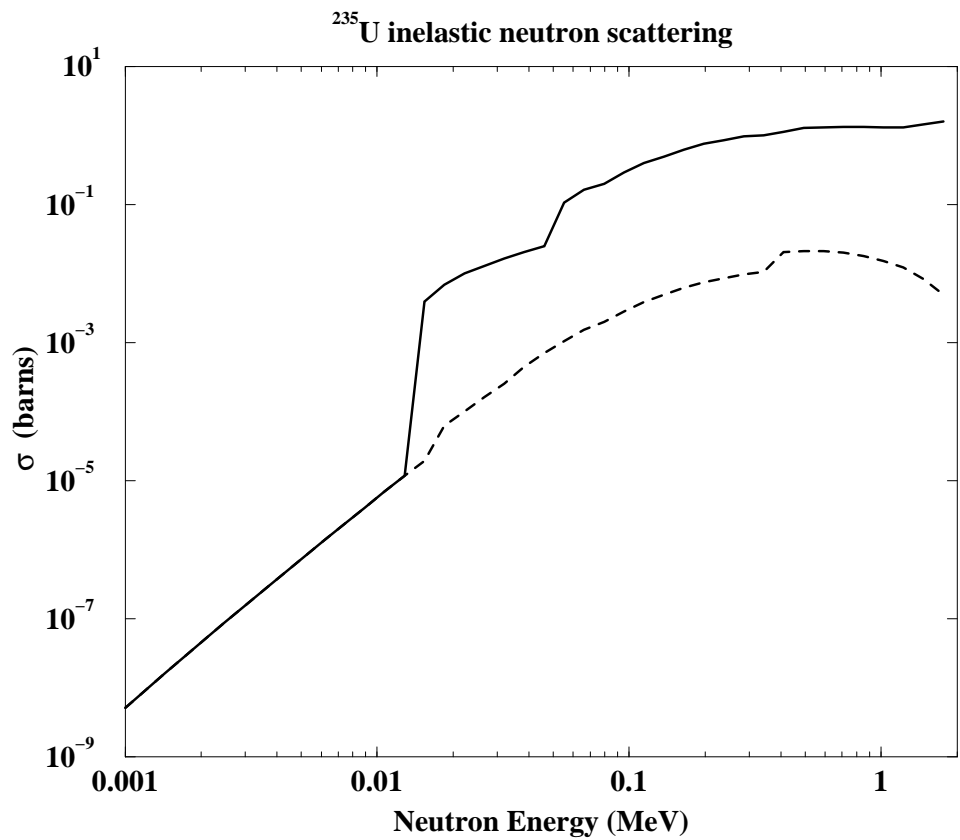}
\caption{The total inelastic cross section of $^{235}$U.
The dotted line shows the predicted cross section for population of the isomer
by the $(n,n')$ reaction.
}
\end{figure}

\begin{figure}
\vspace*{2cm}
\includegraphics[width=5in]{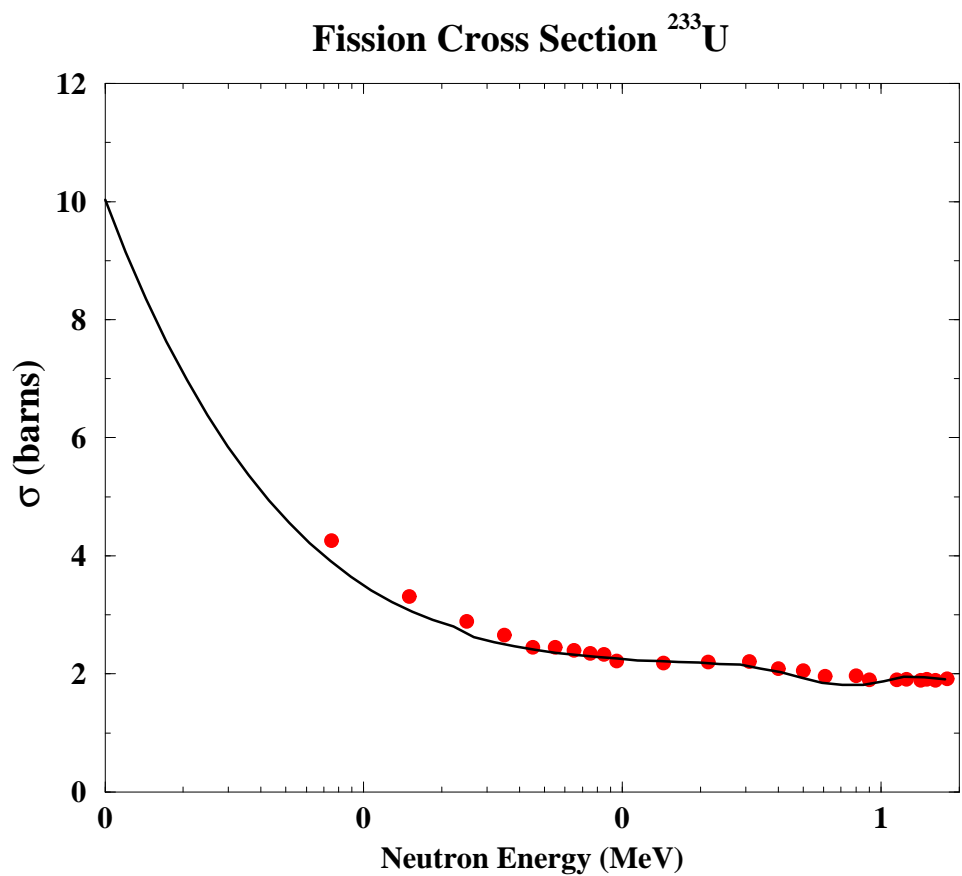}
\caption{The fission cross section of $^{233}$U using the barrier parameters and
transitions states of model 4. The spin and parity of the target nucleus is $5/2^+$.  
}
\end{figure}

\begin{figure}
\vspace*{2cm}
\includegraphics[width=5in]{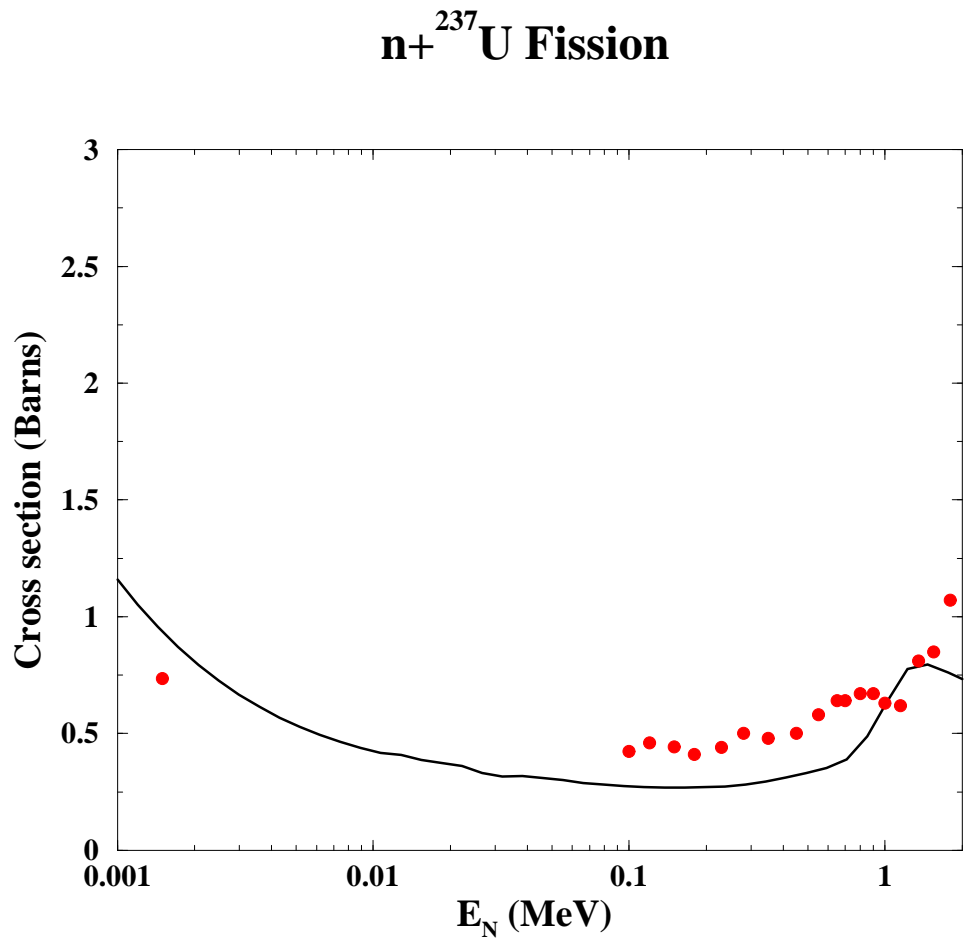}
\caption{The fission cross section of $^{237}$U. The data of ref.\cite{Pommard} are
reduced by a factor 0.63 to agree with critical assembly measurement of
the ratio of $^{237}$U and $^{235}$U fission cross sections\cite{crit}.
}
\end{figure}

\begin{figure}
\vspace*{2cm}
\includegraphics[width=5in]{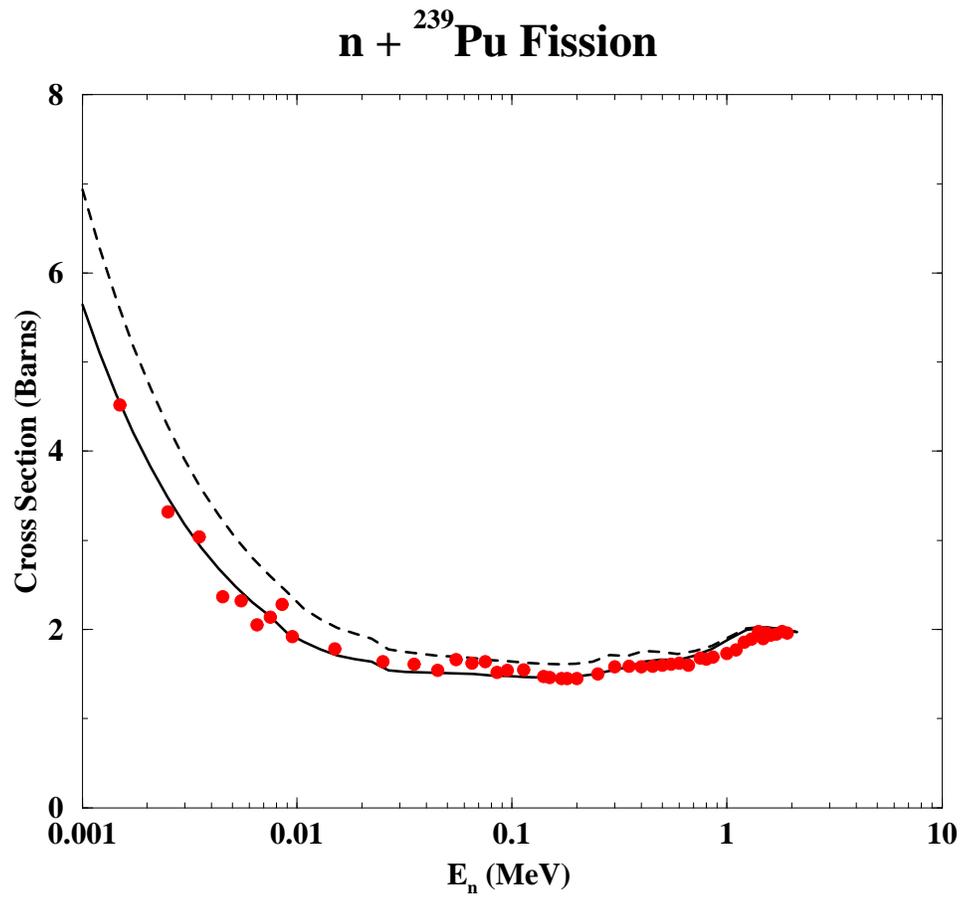}
\caption{The calculated fission cross section of $^{239}$Pu compared
with the experimental data. The solid curve has been calculated from model 4,
while the dashed curve has been calculated with the barrier energy gaps
increased by 0.1 MeV.}
\end{figure}

\begin{figure}
\vspace*{2cm}
\includegraphics[width=5in]{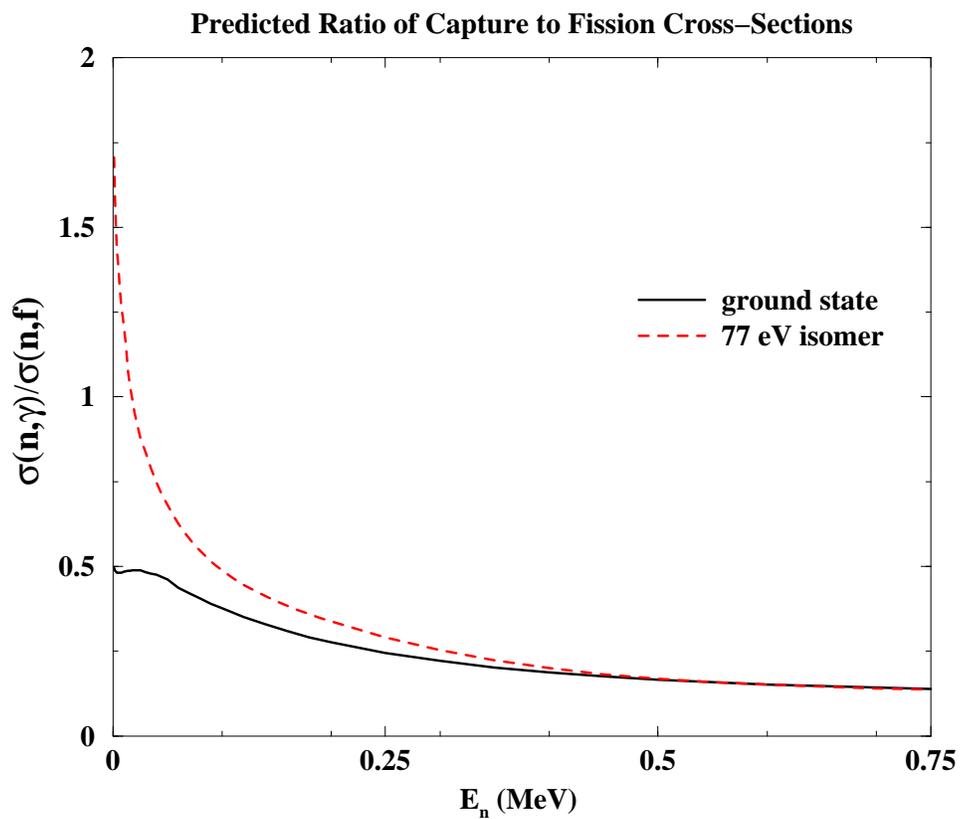}
\caption{The calculated ratio of the neutron capture cross section to the fission cross section
for the isomer and ground state of $^{235}$U. The ratio of these cross sections is
3.4 times larger from the isomer at 1 keV and 2.5 times larger at 10 keV.}
\end{figure}

\end{document}